\documentclass[twocolumn,aps,prb,showpacs,preprintnumbers,amsmath,amssymb]{revtex4}
\usepackage{graphicx}
\usepackage{bm}
\usepackage{gensymb}
\usepackage{color}
\begin{document}
\author{Hena Das$^{1}$, Umesh V. Waghmare$^2$, T. Saha-Dasgupta$^{1}$}
\affiliation{$1$ Department of Materials Science, S.N. Bose National Centre for Basic Sciences,Kolkata 700098, INDIA\\
$2$  Jawaharlal Nehru Centre for Advanced Scientific Research,  Jakkur, Bangalore-560 064, India}
\title{Piezoelectrics by Design: A  Route through Short-period Perovskite Superlattices}
\date{\today}

\begin{abstract}
Using first-principles density functional theory, we study piezoelectricity 
in short-period superlattices made with
combination of ferroelectric and paraelectric components and exhibiting polar discontinuities.
We show that piezoelectric response of such a superlattice can be tuned both in terms of
sign and magnitude with a choice of its components. As these
superlattices with non-switchable polarization do not undergo ferroelectric transitions,
we predict them to exhibit a robust piezoelectric response with weaker temperature dependence compared
to their bulk counterparts. 
\end{abstract}


\maketitle

\par
With advances in experimental techniques, it is now possible to create 
superlattices of various of oxide perovskites (general formula ABO$_3$) 
through layer by layer epitaxial growth. Such superlattices have attracted 
lot of attention in recent years due to the presence of unusual 
properties.\cite{LTO/STO}$^-$\cite{STO/PTO} In particular, these superlattices 
may exhibit polar discontinuities leading to the so-called "polar catastrophe",
a topic of much discussion in recent time in the context of LaAlO$_3$/SrTiO$_3$
thin films.\cite{LS1}$^-$\cite{phase}$^,$\cite{LAO/STO} 

Superlattices with polar discontinuity were studied by Murray and Vanderbilt\cite{Vanderbilt} from first-principles for superlattices consisting of 
ferroelectric (FE) and paraelectric (PE) components, and the "polar catastrophe" was found to 
to result in response of system in terms of polar distortions at the interface.
This was extended in our recent work\cite{HD} to superlattices with ultra-thin 
(1 unit-cell) components of FE and PE layers. Our study\cite{HD} carried out 
on a series of such superlattices constructed from different PE and 
FE components showed that even for ultra-thin superlattices, systems 
containing polar discontinuities have strongly broken inversion symmetry 
and large non-switchable polarization. The magnitudes of the polar distortions 
are as large as those in the corresponding bulk FE materials. Evidently, this 
class of superlattices form promising new material with interesting properties.

In a recent study, tunability of piezoelectric properties of short period 
ferroelectric superlattices with strain was investigated.\cite{Karin} 
Here we demonstrate tunability of robust piezoelectric properties for
superlattices formed of FE and PE layers through (i) polar discontinuity
and (ii) chemical control, using first-principles density functional theory (DFT) based calculations. 

We consider two representative systems: LaAlO$_3$/PbTiO$_3$ (LAO/PTO) and KSbO$_3$/PbTiO$_3$ (KSO/PTO) and design composite superlattices formed out of 
these two superlattices, such as KSbO$_3$/PbTiO$_3$/LaAlO$_3$/PbTiO$_3$ (KSO/PTO/LAO/PTO). The first two systems are composed of alternating 
III-III (A$^{+3}$B$^{+3}$O$_3$)/II-IV (A$^{+2}$B$^{+4}$O$_3$) and 
I-V (A$^{+1}$B$^{+5}$O$_3$)/II-IV perovskite layers respectively, while the 
last system is composed of alternating I-V/II-IV/III-III/II-IV sequence. 

Our calculations are carried out in the plane wave basis within the framework 
of local density approximation (LDA) of DFT using the Vienna Ab-initio simulation 
package (VASP).\cite{vasp1}$^-$\cite{vasp2} The positions of the ions are 
relaxed towards equilibrium until the Hellmann-Feynman forces become less 
than 0.001 eV/$\AA$. A 6$\times$6$\times$4 Monkhorst-Pack k-point mesh and 
450 eV plane-wave cut off are used for the calculations. The polarization is 
calculated using the Berry phase approach\cite{kv} of modern theory of 
polarization. To determine piezoelectric response, strains of
$\pm$ 1\% are applied. The superlattices are constructed along the 
crystallographic {\it z} direction, and the piezoelectric coefficient
relevant to applications is e$_{zz}$, strain is therefore applied along 
the {\it z} direction.

\begin{figure*}
\begin{center}
\includegraphics[scale=0.4]{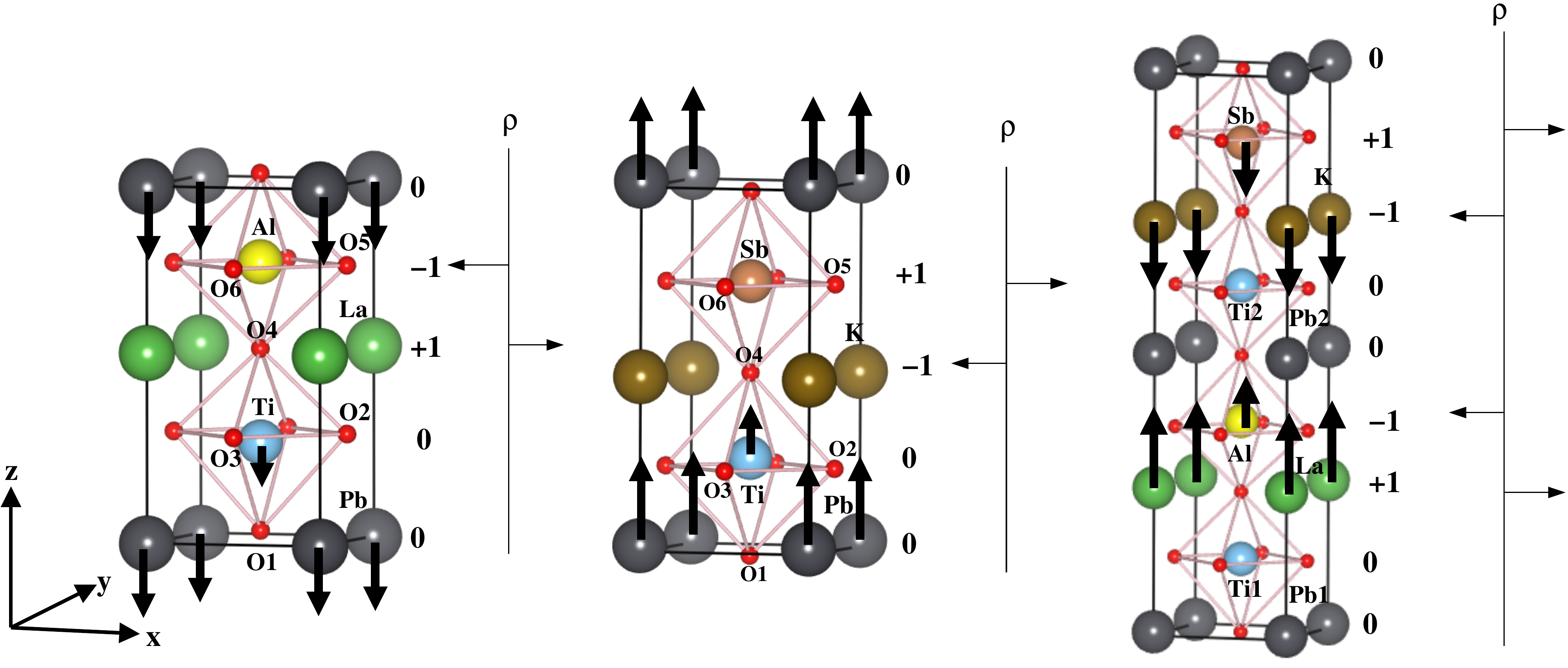} 
\end{center}
\caption{(Color Online) Equilibrium structures of LAO/PTO, KTO/PTO and the composite superlattice KAO/PTO/LAO/PTO (from left to right). The arrows mark the various off-centric movements of the cations measured from the centers of their respective oxygen cages. For clarity, only the off-centric movements larger than 0.05 $\AA$ are shown. The polar discontinuities arising in each superlattices are shown in terms of formal charges ($\rho$).}
\end{figure*}
 
Fig.1 shows equilibrium geometries obtained by relaxation of the atomic 
positions and the out of plane lattice constants (c), with the in-plane 
lattice constant (a) fixed to the LDA bulk lattice constant of PbTiO$_3$ 
in the tetragonal phase. The presence of polar discontinuities drive both 
LAO/PTO and KSO/PTO strongly polar, as found in Ref\cite{HD}. The opposite 
orientation of the polar discontinuity in case of LAO/PTO and KSO/PTO gives 
rise to off-centric movements of Pb and Ti that are oppositely oriented in 
two cases. The off-centric movements driven entirely by the polar 
discontinuity are non-switchable. The third superlattice built up with 
combination of LAO/PTO and KSO/PTO superlattices, impose oppositely directed 
polar fields on the PTO layer, as shown in Fig.1. This leads to the 
interesting situation, where the off-centric movements of Pb and Ti ions 
become negligibly small and the cations in the PE layers move significantly 
with the movements of La and Al being opposite in direction to that of K 
and Sb. The off-centric movements of La, Al, K and Sb measured with respect to the centers of their oxygen cages are found to be as large as 0.25 $\AA$, 
0.16 $\AA$, -0.21 $\AA$ and -0.17 $\AA$ respectively. This may be compared 
with the large displacements of Pb and Ti in case of LAO/PTO and KSO/PTO, 
which are found to be -0.25 $\AA$ and -0.15 $\AA$ respectively for LAO/PTO 
and 0.29 $\AA$ and 0.18 $\AA$ respectively for KSO/PTO. This gives rise to a 
calculated polarization value of 12.79 $\mu C/cm^2$ compared to -50.87 
$\mu C/cm^2$ for LAO/PTO, +57.74 $\mu C/cm^2$ for KSO/PTO, reduced by a 
factor of 4 - 4.5 in composite superlattice compared to its components.

To determine piezoelectric coefficients of the constructed superlattices, in 
the next step we calculate polarization at unstrained structure (P$^{ref}$) 
and at several strained structures (P$^{\epsilon}$), with relaxed atomic 
positions. The slope of the (P$^{\epsilon}$ - P$^{ref}$) versus strain curve 
yields the piezoelectric constants, e$_{zz}$. Fig. 2 shows such curves for 
LAO/PTO, KSO/PTO and KSO/PTO/LAO/PTO superlattices. We find the 
piezoelectric coefficient to be large and negative (-2.58) for LAO/PTO, 
large and positive (+4.08) for KSO/PTO and a small value (0.48) for 
KSO/PTO/LAO/PTO. Using combination of various perovskite materials, we are 
therefore able to design piezoelectrics with e$_{zz}$ of opposite sign and 
varied magnitudes.

\begin{table*}
\caption{Internal displacement gradients as a function of strain ($\partial u_{k,z}$/$\partial \epsilon_{z}$), Born effective charges (Z$_{zz}^*$) for different ions in LAO/PTO and KSO/PTO superlattices. The numbers in the bracket indicate the corresponding numbers for the composite superlattice KSO/PTO/LAO/PTO. Clamped ion contribution to total piezoelectric coefficient (e$^0$), contribution due to internal microscopic strain (e$^1$), the total contribution e$^0$+e$^1$ (e$_{zz}^{indirect}$) and the piezoelectric coefficient obtained from the slopes in Fig.2 (e$_{zz}^{direct}$) are also listed.}
\begin{tabular}{|cccc|cccc|}
\hline
\multicolumn{4}{|c|}{LAO/PTO}&\multicolumn{4}{|c|}{KSO/PTO}\\
\hline
k & $\dfrac{\partial u_{k,z}}{\partial \epsilon _z}$ & Z$_{zz}^*$ & Z$_{zz}^*$ * $\dfrac{\partial u_{k,z}}{\partial \epsilon _z}$ & k & $\dfrac{\partial u_{k,z}}{\partial \epsilon _z}$ & Z$_{zz}^*$ & Z$_{zz}^*$ * $\dfrac{\partial u_{k,z}}{\partial \epsilon _z}$\\ \hline
Pb & -0.15 (0.03) & 3.42 (3.05)& -0.51 (0.08)& Pb & 0.14 (0.03) & 2.88 (3.08)& 0.40 (0.09)\\
La & -0.06 (0.00) & 4.71 (3.70)& -0.28 (0.00)& K & 0.10 (0.02)& 1.23 (1.35)& 0.12 (0.03)\\
Ti & -0.08 (0.02) & 5.56 (4.82)& -0.44 (0.08)& Ti & 0.15 (-0.02)& 5.38 (5.52)& 0.80 (-0.13)\\
Al & -0.04 (-0.06) & 3.69 (3.63)& -0.14 (-0.20)& Sb & 0.12 (0.05)& 6.04 (5.85)& 0.72 (0.31)\\
O1 & 0.13 (0.05) & -3.73 (-3.39)& -0.48 (-0.18)& O1 & -0.14 (-0.03)& -3.11 (-3.43)&0.44 (0.11)\\
O2/O3 & 0.09 (0.00) & -2.11 (-1.74)& -0.19 (0.00)& O2/O3 & -0.09 (-0.02)& -1.61 (-1.68)&0.15 (0.04)\\
O4 & 0.04 (0.06) & -3.58 (-3.36)& -0.14 (-0.18)& O4 & -0.06 (-0.03)& -4.99 (-4.69)&0.30 (0.15)\\
O5/O6 & 0.07 (-0.02)& -2.96 (-2.57)& -0.20 (0.06)& O5 & -0.17 (-0.01)& -2.07 (-2.13)&0.35 (0.02)\\
\hline
e$^0$ (C/m$^2$) & e$^1$ (C/m$^2$) & e$_{zz}^{indirect}$ (C/m$^2$) & e$_{zz}^{direct}$ (C/m$^2$) & e$^0$ (C/m$^2$) & e$^1$ (C/m$^2$) & e$_{zz}^{indirect}$ (C/m$^2$) & e$_{zz}^{direct}$ (C/m$^2$) \\
\hline 
0.17 & -3.01 & -2.85 & -2.58 & -0.10 (0.07) & 4.09 (0.41) & 3.99 (0.45) & 4.08 (0.48)\\
\hline
\end{tabular}
\end{table*}

\begin{figure}
\begin{center}
\includegraphics[scale=0.35]{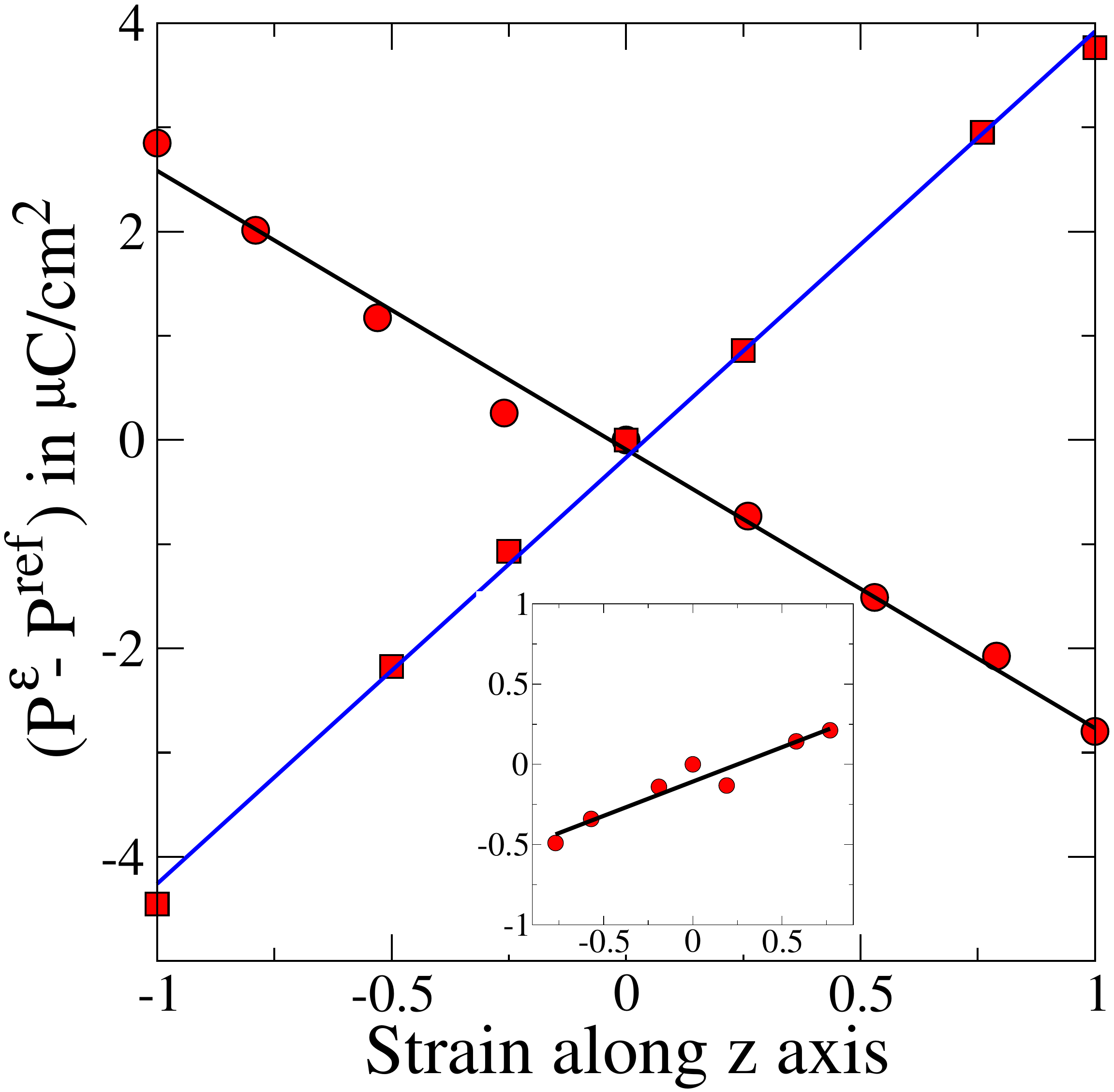} 
\end{center}
\caption{(P$^{\epsilon}$-P$^{ref}$) for LAO/PTO (circles) and KSO/PTO (squares) plotted as function of applied strain along z direction. Inset shows the same plot for KAO/PTO/LAO/PTO. The slope of each curve give the measure of piezoelectric coefficient (see text). Note the change in y-scale between the inset and the main plot.}
\end{figure}

To have large piezoelectric effect, the system must contain ions with large 
Born effective charges (Z$^*$) and they should easily move as a result of 
lattice strain. A piezoelectric coefficient can be separated into two parts: 
a clamped-ion contribution evaluated at vanishing values of the movements of 
the ions ($u$), and a term that is due to the displacements of differently charge 
sublattices\cite{piezo}:
\begin{equation}
e_{zz} = \dfrac{\partial P_z}{\partial \epsilon_{zz}}\Big\vert_{u=0}+\sum_k \dfrac{\partial P_z}{\partial u_k}\dfrac{\partial u_k}{\partial \epsilon_{zz}}
=e^0 + \sum_k \dfrac{q_e}{\Omega} Z^* \dfrac{\partial u_k}{\partial \epsilon_{zz}}
\end{equation}
where $\Omega$ is the volume, $q_e$ is the electronic charge and the subscript $k$ corresponds to the atomic sublattices. 
Table -I lists the various contributions for LAO/PTO, KSO/PTO and KSO/PTO/LAO/PTO superlattices. We find large sensitivity of atomic movements upon strain for Pb, O1 (oxygen at Pb plane) for LAO/PTO system, for Pb, Ti, O1 (oxygen at Pb plane), O5/O6 (oxygens at the Sb plane) for KSO/PTO system. The composite system on the other hand, is far more rigid, with $\partial u$/$\partial \epsilon$ values being order of magnitude smaller. Born effective charges, however, change only slightly between the composite superlattice and the individual superlattices, with the exception of La and O2/O3.

\begin{table}
\caption{The phonon modes that contribute significantly (in the table the percentage contribution is denoted by C) in the piezoelectric response. Frequencies are listed for the equilibrium geometry and that under +1\% and -1\% strain.}
\begin{tabular}{|c|ccc|c|ccc|}
\hline
\multicolumn{4}{|c|}{LAO/PTO}&\multicolumn{4}{|c|}{KSO/PTO}\\
\hline
 & &$\omega$ in cm$^{-1}$& & & & $\omega$ in cm$^{-1}$& \\
 \hline
 C & -1\% & 0 & +1\% & C & -1\% & 0 & +1\% \\
 17 & 371.73 & 363.04 & 352.41 & 10 & 286.24 & 284.69 & 268.81 \\
 10 & 317.19 & 302.56 & 295.93 & 29 & 210.94 & 209.94 & 212.97 \\
 28 & 185.79 & 181.18 & 179.51 & 24 & 200.61 & 194.37 & 196.99 \\
 35 & 116.57 & 115.19 & 110.25 & 28 & 117.44 & 125.83 & 115.37 \\
 \hline
\end{tabular}
\end{table}

Due to non-switchable polarization, it is expected that these superlattices would 
exhibit piezoelectric properties that show weaker temperature dependence 
than the conventional, bulk piezoelectric materials like PbTiO$_3$. To explore this, 
we determine $\Gamma$-point phonons for the superlattices exhibiting 
strong piezoelectric responses, i.e LAO/PTO and KSO/PTO, and decompose 
the displacements of the ions upon strain in terms of these normal modes. Table II lists 
the frequencies of phonon modes that dominantly contribute to the 
piezoelectric coefficients and their changes upon +1\% and -1\% lattice strain.
The frequencies are found to change by 1-4\%, except for the softest ones for 
which the change about 6-8\%. This is in contrast to larger ( 14\% )
change in the soft mode ($\omega$ = 141 cm$^{-1}$) of tetragonal PbTiO$_3$ with
1\% strain, which is known to cause strong temperature dependence in its piezo
and dielectric responses. With relatively smaller Gr\"{u}neisen parameter, 
these superlattices, therefore, should have
a piezoelectric response that is relatively robust as a function of temperature.

In conclusion, we presented the analysis of piezoelectric properties of bi-component, 
ultra-thin superlattices processing polar discontinuities. We demonstrated 
tunability of their piezoelectric response via
appropriate choice of different components, providing a chemical control. 
Moreover their piezoelectric properties are expected to be only weakly
temperature dependent compared to conventional piezoelectrics, opening up a 
new avenue for piezoelectrics by design. 

\par
UVW thanks IBM faculty award and DST for funding. TSD acknowledges support from DST through AMRU project.

\end{document}